\documentclass[runningheads]{llncs}
\usepackage{footmisc}
\usepackage{amssymb}
\usepackage[T1]{fontenc}
\usepackage{amsmath, graphicx, hyperref, multirow, subfigure, caption, booktabs, array, makecell, bm}
\usepackage[hyphenbreaks]{breakurl}
\usepackage{multirow}
\usepackage[T1]{fontenc}
\usepackage{bm}
\usepackage{adjustbox}
\usepackage{enumitem}
\usepackage{graphicx}
\begin{document}
\title{ESTVocoder: An Excitation-Spectral-Transformed Neural Vocoder Conditioned on Mel Spectrogram \thanks{This work was funded by the National Nature Science Foundation of China under Grant 62301521, the Anhui Provincial Natural Science Foundation under Grant 2308085QF200, and the Fundamental Research Funds for the Central Universities under Grant WK2100000033.}}
\titlerunning{ESTVocoder}
\author{Xiao-Hang Jiang \and Hui-Peng Du \and Yang Ai\thanks{Corresponding author.} \and Ye-Xin Lu \and Zhen-Hua Ling}
\authorrunning{X.-H. Jiang et al.}
\institute{National Engineering Research Center of Speech and Language Information Processing, \\University of Science and Technology of China, Hefei, P. R. China \\
 	\email{\{jiang\_xiaohang, redmist\}@mail.ustc.edu.cn}, \email{yangai@ustc.edu.cn}, \email{yxlu0102@mail.ustc.edu.cn}, \email{zhling@ustc.edu.cn}}
%\email{\{yangai, zhling\}@ustc.edu.cn}
\maketitle              % typeset the header of the contribution
\begin{abstract}

This paper proposes ESTVocoder, a novel excitation-spectral-transformed neural vocoder within the framework of source-filter theory. 
The ESTVocoder transforms the amplitude and phase spectra of the excitation into the corresponding speech amplitude and phase spectra using a neural filter whose backbone is ConvNeXt v2 blocks. 
Finally, the speech waveform is reconstructed through the inverse short-time Fourier transform (ISTFT).
The excitation is constructed based on the F0: for voiced segments, it contains full harmonic information, while for unvoiced segments, it is represented by noise. 
The excitation provides the filter with prior knowledge of the amplitude and phase patterns, expecting to reduce the modeling difficulty compared to conventional neural vocoders. 
To ensure the fidelity of the synthesized speech, an adversarial training strategy is applied to ESTVocoder with multi-scale and multi-resolution discriminators. 
Analysis-synthesis and text-to-speech experiments both confirm that our proposed ESTVocoder outperforms or is comparable to other baseline neural vocoders, e.g., HiFi-GAN, SiFi-GAN, and Vocos, in terms of synthesized speech quality, with a reasonable model complexity and generation speed. 
Additional analysis experiments also demonstrate that the introduced excitation effectively accelerates the model's convergence process, thanks to the speech spectral prior information contained in the excitation.

\end{abstract}

\section{Introduction}
Text-to-speech (TTS) is a technique that utilizes machines to convert written text content into audible speech. 
With the recent advancements in deep learning, there has been a significant improvement in the clarity and naturalness of synthesized speech.
A TTS system typically consists of two key components,i.e., an acoustic model and a vocoder. 
The acoustic model is responsible for predicting acoustic features (e.g., mel spectrogram) from the input text. 
Subsequently, the vocoder uses these features to generate the final speech waveform.
Besides TTS, other fields such as voice conversion (VC) and singing voice synthesis (SVS) also require a vocoder to reconstruct waveform. 
Thus, a robust vocoder is crucial for the field of speech signal processing, which is also the focus of this paper.
%, so a powerful vocoder can ensure the generation of high-quality speech. In recent years, many studies have been conducted to improve the performance of vocoders.

Traditional vocoders, such as STRAIGHT \cite{kawahara1999restructuring} and WORLD \cite{morise2016world}, synthesize speech waveform using traditional signal processing methods.
Although they are computationally simple and fast, result in synthesized speech with lower naturalness. 
With the development of deep learning, the WaveNet \cite{oord2016wavenet} represents a significant milestone in synthesized speech quality. 
WaveNet is an autoregressive neural vocoder capable of producing natural and clear speech waveforms. 
Unlike traditional vocoders that rely on traditional signal processing methods, WaveNet relies entirely on end-to-end neural network training. 
%Since then, neural vocoders can essentially be divided into two main categories: autoregressive models and non-autoregressive models. 
However, autoregressive neural vocoders generate audio samples sequentially and use previously generated samples to create new ones, leading to extremely low efficiency and high computational complexity. 
To address the issue of low efficiency in autoregressive models, researchers have proposed various alternative approaches, including knowledge distillation-based models \cite{oord2018parallel,ping2018clarinet}, flow-based models \cite{prenger2019waveglow,ping2020waveflow}, and glottis-based models \cite{juvela2019glotnet,valin2019lpcnet}. 
Although these models have improved inference efficiency, their overall computational complexity remains high.
% For many years, traditional vocoders such as WORLD and STRAIGHT have been widely used. But they also have significant drawbacks, such as low naturalness in the synthesized speech. The WaveNet model, introduced in 2016, marked an important milestone in the history of speech synthesis. It was the first to exhibit the deep learning model's ability to generate high-quality speech directly from raw audio waveforms. WaveNet is an autoregressive model based on neural networks that can produce speech waveforms with extremely high naturalness and clarity. Unlike traditional parametric voice codecs that incorporate prior knowledge about the audio signal, WaveNet relies solely on end-to-end training. Since the emergence of WaveNet, neural network vocoders can be basically divided into two categories: autoregressive models and non autoregressive models. Autoregressive models, such as WaveNet, generate audio samples in order and use previously generated samples to generate new ones, which inevitably leads to lower efficiency, requiring a lot of computational power and slow generation speed. To address the issue of low efficiency in autoregressive models, researchers have proposed various alternative methods, including knowledge distilling based models, flow based models, and glottis based models. Although these models greatly improve inference efficiency, their overall computational complexity remains high. 

To overcome the aforementioned issues, non-autoregressive neural vocoders have gradually been proposed. 
Non-autoregressive models generate all samples in parallel, offering high computational efficiency. 
For instance, HiFi-GAN \cite{kong2020hifi} vocoder maintains high naturalness in synthesized audio thanks to the generative adversarial network (GAN) \cite{goodfellow2014generative} based training while balancing high generation speed.
% For instance, vocoders based on generative adversarial network (GAN) \cite{goodfellow2014generative} such as HiFi-GAN \cite{kong2020hifi}, has been proposed. Compared to autoregressive models, HiFi-GAN maintains high naturalness in synthesized audio while balancing high generation speed. 
However, there is still room for efficiency improvement with these vocoders, as they directly predict high-temporal-resolution waveforms from input acoustic features. 
The substantial discrepancy in time resolution between the acoustic fratures and waveforms, results in extensive upsampling operations on the acoustic features, leading to significant computational demands.
%, the Neural Source Filter (NSF) model \cite{wang2019neural} synthesizes speech based on explicit fundamental frequency (F0) and mel spectrogram features. Additionally,
% On the other hand, non autoregressive models generate all samples independently, and the process is parallelized, resulting in high computational efficiency, which has attracted the attention of researchers. The Neural Source Filter (NSF) model integrates speech production processes with neural networks, enabling the accurate forecasting of speech audio waveforms based on explicit fundamental frequency (F0) and mel spectrogram features. In addition, vocoders based on generative adversarial networks have also been proposed, among which HiFi-GAN has achieved excellent performance. Compared with past autoregressive models, HiFi-GAN maintains high naturalness in synthesized audio while balancing high computational and generation efficiency. However, these vocoders based on Generative Adversarial Networks (GANs) still have room for improvement, as they directly predict waveforms from input acoustic features. The discrepancy in time resolution between these two is substantial, necessitating the use of multiple transposed convolutions to upsample the acoustic features. This process remains computationally intensive. 
Thus, subsequent neural vocoders have adopted the approach of predicting amplitude and phase spectrum and then reconstructing the waveform using the inverse short-time Fourier transform (ISTFT). 
For instance, Vocos \cite{siuzdak2023vocos} with ConvNeXt blocks \cite{liu2022convnet} as backbone, directly predicts the amplitude and phase spectrum at the same temporal resolution from the input acoustic features, thereby maintaining the same feature resolution at all frame rates. 
Vocos has increased its generation speed by more than tenfold compared to HiFi-GAN while maintaining high-quality synthesized speech. 
%Additionally, there are vocoders that combine source filter models with Generative Adversarial Networks, such as SiFi-GAN \cite{yoneyama2023source}, which introduces source filter auxiliary modeling into HiFi-GAN.

Most of the aforementioned vocoders only use the mel spectrogram as input, which is convenient, but the mel spectrogram is a compressed representation of the amplitude spectrum and may lose some acoustic details. 
Therefore, many vocoders that utilize other acoustic features have also been proposed. 
A common approach is to enhance vocoder performance based on the source-filter theory framework by introducing the F0 as an additional acoustic feature.
Neural source filter (NSF) model \cite{wang2019neural} is a pioneer in applying neural networks within the source-filter framework, which synthesizes speech waveform directly based on explicit F0 and mel spectrograms. 
Recently, some works combining source-filter vocoders with GAN-based training, such as SiFi-GAN \cite{yoneyama2023source} and SF-GAN \cite{lu2022source}, have been proposed. 
This type of vocoders generates excitation based on the F0. 
The excitation waveform is then processed through a neural filter conditioned on the mel spectrogram to directly produce the final speech waveform. 
Experiments have shown that after introducing the additional F0 features, the quality of the speech generated by these vocoders is obviously improved. 
However, their excitations are often constructed based on single F0, lacking harmonic information, which may impact the reconstruction performance of the neural filter.
In addition, these methods often rely on direct transformation of the excitation waveform, still leaving room for improvements in efficiency and model complexity. 
Excitation-spectral-transformed methods in source-fileter-based neural vocoders have not yet been thoroughly investigated.

To achieve high-fidelity speech synthesis as well as efficient training and rapid generation speed, we propose a novel excitation-spectral-transformed neural vocoder called ESTVocoder. 
The ESTVocoder is designed based on the source-filter theory and first produces an excitation according to the F0.
%Compared to the original Vocos, ESTVocoder utilizes the mel spectrogram to guide the excitation signal based on fundamental frequency (F0) to transform into the final waveform.
Compared to the single-F0-based excitation used by SiFi-GAN \cite{yoneyama2023source} and SF-GAN \cite{lu2022source}, the proposed ESTVocoder utilizes a full-harmonic excitation which contains richer spectral information. 
Subsequently, a neural filter with ConvNeXt v2 blocks \cite{woo2023convnext} as the backbone transforms the amplitude and phase spectra of the excitation into the corresponding amplitude and phase spectra of the speech, conditioned on the mel spectrogram. 
Finally, the speech waveform is reconstructed via ISTFT. 
Both analysis-synthesis and TTS experimental results show that our proposed ESTVocoder outperforms or is comparable to HiFi-GAN, SiFi-GAN, and Vocos, in terms of synthesized speech quality. 
Our proposed ESTVocoder also has an extremely fast training convergence speed, thanks to the introduction of spectral prior information contained in the excitation. 

This paper is organized as follows:
In Section \ref{sec:pagestyle}, we provide details on our proposed ESTVocoder.
In Section \ref{sec:exp}, we present our experimental results.
Finally, we give conclusions in Section \ref{sec:con}.

\begin{figure}[t]
	\centering
	\includegraphics[width=1\textwidth]{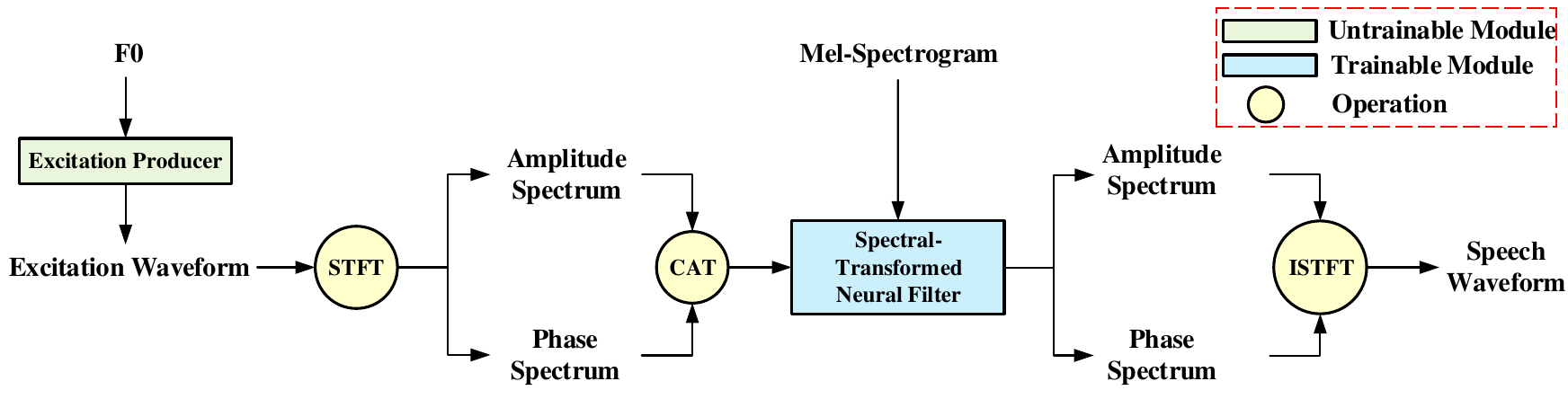}
	\caption{Overview of the proposed ESTVocoder. Here, \emph{CAT} represents the concatenation operation.} \label{Overview}
\end{figure}

\section{Proposed Method}
\label{sec:pagestyle}

\subsection{Overview}

Fig. \ref{Overview} shows an overview of the proposed ESTVocoder model. 
In ESTVocoder, the F0 sequence $\bm{f}\in\mathbb{R}^F$ first passes through an excitation producer to produce an excitation waveform $\bm{e}\in\mathbb{R}^T$, i.e.,
\begin{align}
\label{equ: e_producer}
\bm{e}=EP(\bm{f},S_r),
\end{align}
where $EP$ represents the excitation producer and $S_r$ is waveform sampling rate.
$F$ and $T$ denote the number of frames of the F0 and the number of excitation waveform samples, respectively. 
Assuming the frame shift for extracting F0 and spectral features from the waveform is $w_s$, it holds $T=F\cdot w_s$. 
Then, a spectral-transformed neural filter transforms the amplitude spectrum $\bm{A}_e\in\mathbb{R}^{F\times N}$ and phase spectrum $\bm{P}_e\in\mathbb{R}^{F\times N}$ extracted from the excitation waveform $\bm{e}$ by STFT into the corresponding speech amplitude spectrum $\hat{\bm{A}}\in\mathbb{R}^{F\times N}$ and phase spectrum $\hat{\bm{P}}\in\mathbb{R}^{F\times N}$, with the mel spectrogram $\bm{M}\in\mathbb{R}^{F\times M}$ as conditions, i.e.,
\begin{align}
\label{equ: e}
\bm{A_e},\bm{P_e}=STFT(\bm{e}),
\end{align}
\begin{align}
\label{equ: S}
\bm{\hat{A}},\bm{\hat{P}}=ST\text{-}NF(\bm{A_e},\bm{P_e} \vert \bm{M}),
\end{align}
where $ST\text{-}NF$ denotes the spectral-transformed neural filter. $N$ and $M$ represent the number of the frequency bins of the amplitude/phase spectra and mel spectrogram, respectively. 
Finally, the predicted speech amplitude and phase spectra are reconstructed into the speech waveform $\hat{\bm{x}}\in\mathbb{R}^{T}$ through the ISTFT, i.e.,
\begin{align}
\label{equ: ISTFT}
\hat{\bm{x}}=ISTFT(\bm{\hat{A}} \cdot exp(j\bm{\hat{P}})).
\end{align}

\subsection{Excitation Producer}
\label{sec:excit}

The excitation producer generates the corresponding excitation waveform $\bm{e}$ based on the frame-level F0 sequence $\bm{f}$. 
First, a point-level F0 sequence $\bm{f}_{pl} = \left[f_1, \ldots, f_T\right]^\top$ is first generated by repeating the F0 value of $\bm{f}$ within each frame, which then serves as the input to the excitation producer. 
The excitation producer outputs an excitation signal $\bm{e}=\left[e_1, \ldots, e_T\right]^\top$, which is a sine-based signal at voiced segments and Gaussian white noise at unvoiced segments. 
Specifically, the excitation waveform can be represented as:

\begin{equation}
\label{eq1}
e_t=\left\{\begin{array}{ll}\sum\limits_{k=1}\limits^K\alpha \sin(\sum\limits_{h=1}\limits^t 2\pi k\frac{f_h}{S_r} ) + n_t,&f_t > 0,t\in V_j\\ \frac{1}{3\sigma}n_t,&f_t=0\end{array}\right.,
\end{equation}
where $K = \left\lfloor \frac{S_r/2}{\text{min}(\bm{f})} \right\rfloor$ represents the minimum number of harmonics required to cover the entire frequency band. $f_t = 0$ indicates that the $t$-th sampling point is part of an unvoiced frame, $n_t \sim \mathcal{N}(0, \sigma^2)$ is a Gaussian noise, $V_j$ represents the $j$-th voiced segment to which the $t$-th sampling point belongs, $\alpha$ and $\sigma$ are hyperparameters. 

\begin{figure}[t]
	\centering
	\includegraphics[width=0.8\textwidth]{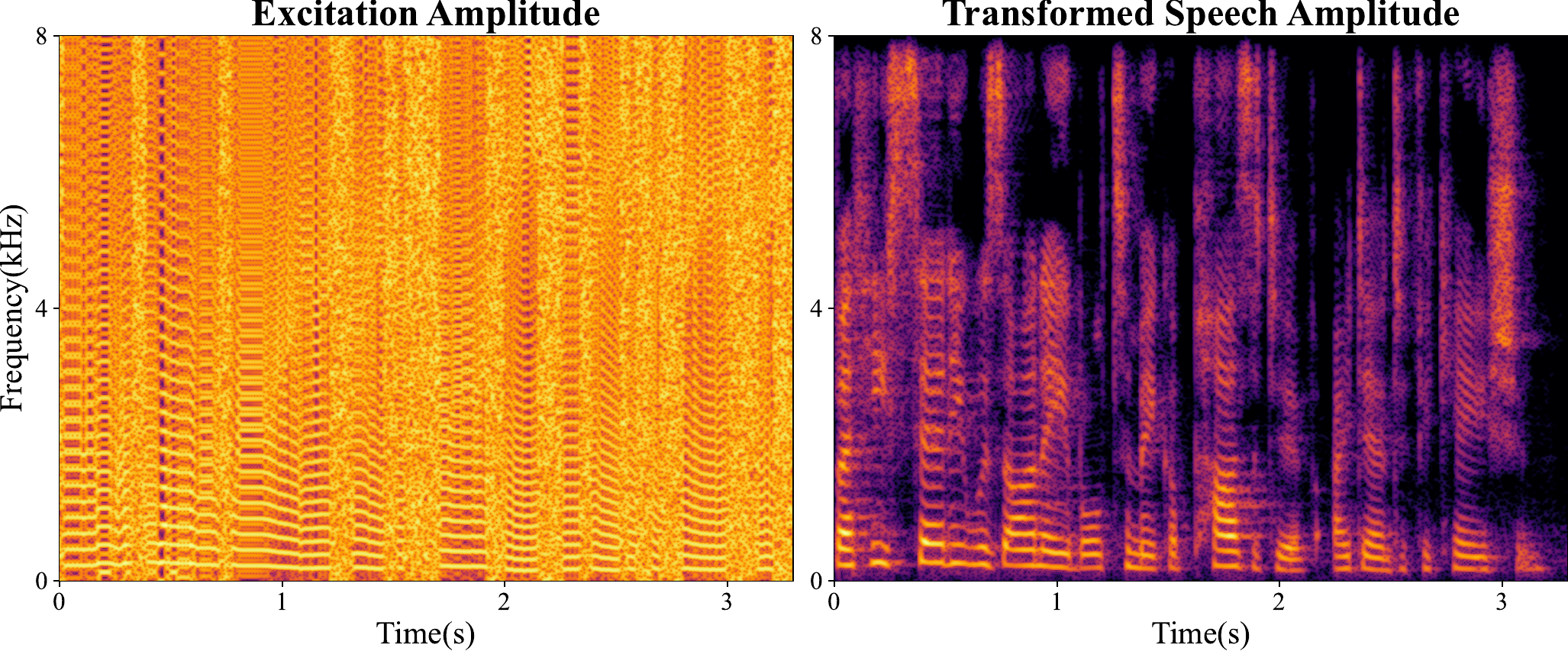}
	\caption{Visualization of the amplitude spectrum of the excitation waveform and the speech waveform generated by ESTVocoder. } \label{analysis1}
\end{figure}

The left subplot of Fig. \ref{analysis1} gives a visualization example of the amplitude spectrum of the excitation waveform. 
Compared to previous works \cite{lu2022source,ai2020neural}, the most significant difference in our proposed excitation is that our excitation includes all possible harmonic information. 
In this way, the produced excitation already includes the basic acoustic pattern of the speech waveform, expecting to effectively alleviate the modeling and learning difficulty for the subsequent neural filter. 
% The amplitude spectrum of the resulting waveform already has the basic shape of a speech amplitude spectrum. On this basis, guided by the mel spectrogram, convergence can be achieved with fewer steps. We believe that such an excitation signal can reduce the learning difficulty of the model, thereby accelerating the model's convergence speed, while also diminishing the smooth artificial artifacts in the amplitude spectrum of synthesized speech. 

% where $v_t = 0$ or $1$ denotes that $e_t$ belongs to unvoiced or voiced segment, $\alpha$ and $\sigma$ are hyperparameters, $g(\cdot)$ represents a DNN-based transformation, $n_t \sim \mathcal{N}(0, \sigma^2)$ is a Gaussian noise, $\phi \in (-\pi, \pi]$ is a random initial phase, and $N_s$ denotes the waveform sampling rate. In the source module, only the DNN is trainable.

% As shown in Fig. \ref{analysis2}, the excitation signal is transformed into the final waveform under the guidance of the mel spectrogram, and it can be seen that the excitation signal includes all the harmonics in the final waveform. For example, at the 1-second mark, the effect of ESTVocoder on the spectrum can be observed, from left to right, from shapeless to shaped.

\subsection{Spectral-Transformed Neural Filter}
\label{ssec:block}

Unlike the waveform-transformed neural filters used in other vocoders, the neural filter in ESTVocoder transforms the amplitude spectrum $\bm{A}_e$ and phase spectrum $\bm{P}_e$ of the excitation waveform $\bm{e}$ into the speech amplitude spectrum $\hat{\bm{A}}$ and phase spectrum $\hat{\bm{P}}$ conditioned on the mel spectrogram $\bm{M}$, and then reconstructs the speech waveform $\hat{\bm{x}}$ through ISTFT. 
Fig. \ref{analysis1} provides an visualization example of amplitude spectral transformation (from the left subplot to the right subplot). 
Therefore, the role of the neural filter is to achieve the gradual spectral refinement using a neural network, guided by the information from the mel spectrogram. 
As mentioned in Section \ref{sec:excit}, the excitation spectrum already has the rudiments of the speech spectrum, essentially providing the neural filter with prior spectral information. 
We expect that such an excitation can reduce the learning difficulty of the neural filter model, thereby accelerating the model's convergence speed, while also diminishing the smooth artificial artifacts in the amplitude and phase spectra of the synthesized speech. 

Specifically, the amplitude and phase spectra of the excitation are concatenated along the dimensional axis and then passed through a dimension-reduction layer. 
The mel-spectrogram, used as a condition, is first injected into a dimension-expansion layer to match the dimension of the dimension-reduced amplitude-phase feature. 
%The mel-spectrogram, used as a condition, is concatenated with this dimension-reduced feature and then passed through another dimension-reduction layer. 
Then, these two features with same dimension are added together and passed through the backbone network (i.e., ConvNeXt v2) along with a layer normalization layer and a feed-forward layer to predict the speech amplitude and phase spectra. 
Inspired by Vocos \cite{siuzdak2023vocos}, we split the output of the above feed-forward layer into two features along the dimensional axis. One feature generates the amplitude spectrum through exponential activation, while the other feature generates the phase spectrum by first applying sine/cosine calculations and then the atan2 function. 

% ESTVocoder features two types of dimension-reduction convolutional layers. One type of dimension-reduction convolutional layer serves to reduce the dimensions of the amplitude and phase spectrum of the excitation signal after STFT transformation. The other type of dimension-reduction convolutional layer is for reducing the dimensions of the concatenated amplitude and phase spectrum of the excitation signal along with the input mel spectrogram to fit the input requirements of the ConvNeXt v2 network. The low-dimensional concatenated amplitude and phase of the excitation signal is input into the ConvNeXt v2 network, while the transformed mel spectrogram is also input as a condition.

\begin{figure}[t]
	\centering
	\includegraphics[width=1\textwidth]{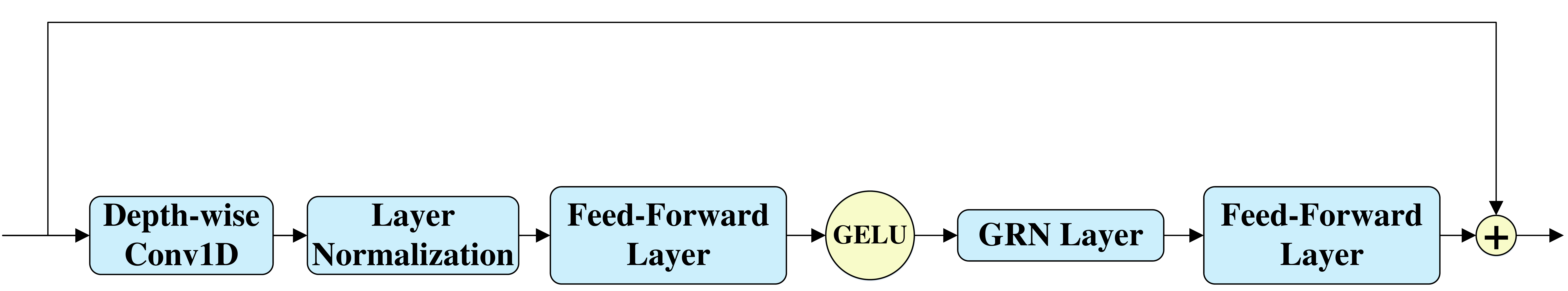}
	\caption{The specific compositional structure of the ConvNeXt v2 block.} \label{ConvNeXtv2}
\end{figure}

The neural filter adopts a ConvNeXt v2 \cite{woo2023convnext} as its backbone, which has been proven to possess superior modeling capabilities compared to the commonly used ResNet networks \cite{he2016deep} in HiFi-GAN \cite{kong2020hifi} and the ConvNeXt network \cite{liu2022convnet} in Vocos \cite{siuzdak2023vocos}. 
The backbone ConvNeXt v2 network consists of multiple cascaded identical ConvNeXt v2 blocks.  
As shown in Fig. \ref{ConvNeXtv2}, the key components of the ConvNeXt v2 block include a 1D depth convolutional layer, a layer normalization layer, a feed-forward layer that projects features to higher dimensions, a Gaussian Error Linear Unit (GELU) activation \cite{hendrycks2016gaussian}, a global response normalization (GRN) layer \cite{woo2023convnext}, and an another feed-forward layer that projects features back to their original dimensions. 
The GRN layer comprises three integral steps, i.e., aggregating global features, normalizing these features, and calibrating them, thereby enhancing feature diversity and ultimately boosting the representational quality. 
The final output of the ConvNeXt v2 block is obtained by adding the input of the 1D depth convolutional layer and the output of the last feed-forward layer (i.e., residual connection).

% The output of the ConvNeXt v2 network is then passed through a normalization layer and a fully connected layer before being split into amplitude and phase components. Subsequently, a complex number $S$ is constructed, and an inverse STFT is applied to $S$ to obtain the final audio.

% Therefore, the process of waveform reconstruction by the vocoder using the excitation signal can be represented by the following formula:
% \begin{align}
% \label{equ: e}
% \bm{A_e},\bm{P_e}=STFT(\bm{e}),
% \end{align}
% \begin{align}
% \label{equ: S}
% \bm{\hat{A}},\bm{\hat{P}}=E\text{-}Vocos(\bm{A_e},\bm{P_e} \vert \bm{M})
% \end{align}
% \begin{align}
% \label{equ: ISTFT}
% \hat{\bm{x}}=ISTFT(\bm{\hat{A}} \cdot exp(j\bm{\hat{P}})),
% \end{align}
% where $\bm{A_e},\bm{P_e}$ and $\bm{\hat{A}},\bm{\hat{P}}$ represent the amplitude and phase spectrum of the excitation signal and generated waveform $\hat{\bm{x}}$. $\bm{M}$ is the mel spectrogram.

\subsection{Training Criteria}
\label{ssec:loss}

In ESTVocoder, only the neural network filter contains trainable parameters, and it is trained using an adversarial training approach. 
The ESTVocoder uses multi-period discriminator (MPD) \cite{kong2020hifi} and multi-resolution discriminator (MRD) \cite{jang2021univnet} for adversarial training to ensure the quality of synthesized waveforms. 
\begin{itemize}[nosep, leftmargin=*]

\item {}{\textbf{MPD}}: The MPD consists of 5 parallel sub-MPDs that operate in parallel. 
Each sub-MPD converts the input synthesized waveform $\hat{\bm{x}}$ or natural waveform $\bm{x}$ into a 2D periodic map based on the preset period. 
This mapping result then goes through a series of 5 consecutive processing stages, each stage containing a layer of 2D convolution and a leaky rectified linear unit (LReLU) activation function. 
The final output of the processing chain is further processed through a 2D convolutional output layer to generate discriminative scores. The period parameters are set to 2, 3, 5, 7, and 11, respectively. 

\item {}{\textbf{MRD}}: The MRD is configured with 3 parallel sub-MRDs. 
Each sub-MRD uses the amplitude spectrum extracted from $\hat{\bm{x}}$ or $\bm{x}$ as input, based on specific STFT parameters. 
Assuming the STFT parameters for extracting the input amplitude and phase spectrum for the neural filter are: [frame length, frame shift, FFT point number] = [$w_l$, $w_s$, $2N+1$].
We set the STFT parameters for the three sub-MRDs as [$w_l/2$, $w_s/2$, $(2N+1)/2$], [$w_l$, $w_s$, $2N+1$] and [$2w_l$, $2w_s$, $2(2N+1)$], respectively. 
Each sub-MRD consists of blocks made up of 2D convolutional layer and LReLU activation, followed by a 2D convolutional layer that provides the final output. 

\end{itemize}

We use the hinge form of adversarial loss. 
Assuming $D*$ represents the sub-discriminator of MPD or MRD, the corresponding generator adversarial loss and discriminator adversarial loss are as follows.
\begin{align}
\label{equ: adv G}
\mathcal L_{adv-G}^*=\mathbb{E}_{\hat{\bm{x}}}\max \left(0,1-D^*(\hat{\bm{x}})\right)\text{,}
\end{align}
\begin{align}
\label{equ: adv D}
\mathcal L_{adv-D}^*=\mathbb{E}_{\left(\hat{\bm{x}},\bm{x}\right)}\left[\max \left(0,1-D^*(\bm{x})\right) + \max \left(0,1+D^*(\hat{\bm{x}})\right)\right]\text{.}
\end{align}
Additionally, feature matching loss $\mathcal L_{FM}^*$ \cite{kumar2019melgan} is also utilized, characterized by the summation of the mean absolute error (MAE) between the corresponding intermediate layer outputs of sub-discriminator $D^*$ when provided with inputs $\hat{\bm{x}}$ or $\bm{x}$.
% In ESTVocoder, end-to-end training is achieved through mel spectrogram reconstruction loss function, generative adversarial loss function, and feature matching loss function.

In addition, we also incorporate the mean absolute error (MAE) loss on the mel spectrogram between the extracted mel spectrogram $\hat{\bm{M}}$ and natural one $\bm{M}$ derived from synthesized waveform $\hat{\bm{x}}$ and natural one $\bm{x}$, respectively, i.e.,

\begin{align}
\label{equ: M Loss}
\mathcal L_{M}=\dfrac{1}{FM}\cdot\mathbb{E}_{\left(\hat{\bm{M}},\bm{M}\right)}\left( \left\lVert \hat{\bm{M}}-\bm{M}\right\rVert_1\right)\text{.}
\end{align}
Therefore, the final GAN-based losses for generator (i.e., the neural filter) and discriminator (i.e., the MPD and MRD) are respectively defined by the following expressions.
\begin{align}
\label{equ: GAN G}
\mathcal L_{G}=\sum_{i=1}^5 \left(\mathcal L_{adv-G}^{Pi}+\mathcal L_{FM}^{Pi} \right)+\lambda_{MRD}\sum_{j=1}^3 \left(\mathcal L_{adv-G}^{Rj}+\mathcal L_{FM}^{Rj} \right)+\lambda_{M}\mathcal L_{M}\text{,}
\end{align}
\begin{align}
\label{equ: GAN D}
\mathcal L_{D}=\sum_{i=1}^5 \mathcal L_{adv-D}^{Pi}+\lambda_{MRD}\sum_{j=1}^3 \mathcal L_{adv-D}^{Rj}\text{,}
\end{align}
where $Pi$ and $Rj$ represent $i$-th sub-MPD and $j$-th sub-MRD, $\lambda_{MRD}$ and $\lambda_{M}$ are hyperparameters.
% The final generator loss is a linear combination of the aforementioned mel spectrogram loss and GAN-based loss, i.e.,
% \begin{align}
% \label{equ: G_final}
% \mathcal L = \lambda_{M}\mathcal L_{M}+\mathcal L_{G}\text{ ,}
% \end{align}
% where $\lambda_{M}$ are hyperparameters.
The training process follows the standard training mode of GAN, i.e., using $\mathcal L_G$ and $\mathcal L_{D}$ to train the generator and discriminator alternately.

\subsection{Analysis-synthesis and TTS applications}

During the test stage, we apply the proposed ESTVocoder on two tasks, i.e., the analysis-synthesis task and TTS task. 
The process of analysis-synthesis applications is consistent with the training process. 
We first extract the natural F0 and the natural mel spectrogram from the natural speech and then inject them into the well-trained ESTVocoder to generate the speech waveform.

% In the analysis-synthesis (AS) task, we first extract the F0 from the audio. Subsequently, we utilize the source module to generate an excitation signal from the F0. The amplitude and phase spectrum of this excitation signal are obtained through Short-Time Fourier Transform (STFT). These components are concatenated and, along with the mel spectrogram, are fed into the modified ConvNeXt v2 network \cite{woo2023convnext} for further processing. 

\begin{figure}[t]
	\centering
	\includegraphics[width=1\textwidth]{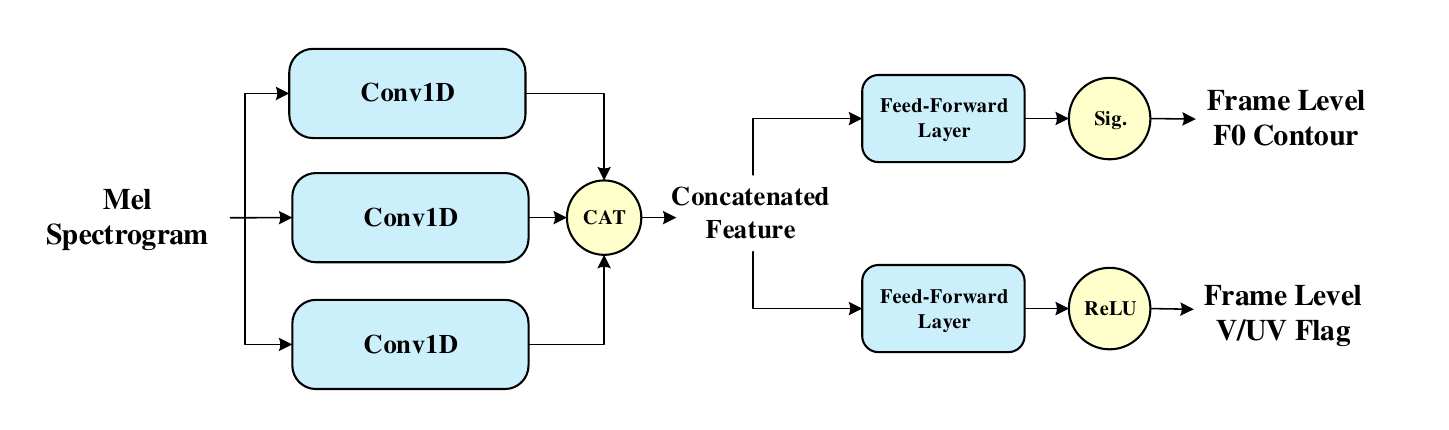}
	\caption{The structure of the F0 predictor. Here, \emph{CAT} and \emph{Sig.} represents the concatenation operation and sigmoid activation.} \label{f0predictor}
\end{figure}

In TTS tasks, after predicting the mel spectrogram from text via an acoustic model, we use an F0 predictor \cite{lu2022source} to predict the F0 from the mel spectrograms. 
Then the predicted F0 and mel spectrogram are inject into the well-trained ESTVocoder to generate the speech waveform.
%Beyond this step, the process remains consistent with the analysis-by-synthesis approach.
The structure of the F0 predictor is shown in Fig. \ref{f0predictor}. 
The mel spectrogram is first input into three parallel convolutional layers with different kernel sizes to obtain three sets of intermediate features. 
Subsequently, the concatenated features are inputted into two parallel feed-forward layers with different activation functions (Sigmoid and ReLU, respectively) to separately derive the F0 contour and the voiced/unvoiced (V/UV) flag. 
Finally, multiply the F0 contour and the V/UV flag to obtain the final F0 sequence.

\section{Experiments}
\label{sec:exp}

\subsection{Experimental Setup}
\label{ssec:expset}
\subsubsection{Dataset and Implementation.} 
We conducted experiments on the LJSpeech \cite{ito2017lj} dataset, which included 13,100 speech utterances from an English female for a total of approximately 24 hours. 
All speech waveforms were downsampled to 16 kHz for our experiments. 
We randomly selected 10,480 utterances for training, 1,310 for validation, and remaining 1,310 for testing. 
When performing the STFT, the frame length was 640 (i.e., $w_l=640$), the frame shift was 160 (i.e., $w_s=160$), and the number of FFT points was 1024 (i.e., $N=513$). 

In the neural filter model, before the ConvNeXt v2 block, both the layer dimension-reduction and dimension-expansion layers had 512 nodes. 
The number of ConvNeXt v2 blocks in the ConvNeXt v2 network was set to 8.
For each ConvNeXt v2 block, the kernel size and channel size were set to 7 and 512, respectively. 
The model was trained on on a single NVIDIA RTX TITAN GPU and optimized using the AdamW optimizer with $\beta_1$ = 0.8, $\beta_2$ = 0.99, and weight decay of 0.01. 
The learning rate was set initially to $2\times10^{-4}$ and scheduled to decay with a factor of 0.999 at every epoch.

% \subsubsection{Implementation.}\label{above}
% In the proposed APCodec+, the number of ConvNeXt v2 blocks in each backbone network was set to 8.
% For each ConvNeXt v2 block the kernel size and channel size were set to 7 and 512, respectively. The dimensionality of the quantizer codebook is set to 32. The dimensionality used to extract the mel spectrogram is set to 80 for computing the spectral-level loss.

% We trained our proposed APCodec+ up to 1 million steps on VCTK for pre-training, 0.5 million steps for stage 1 and 1 million steps for stage 2 on Expresso dataset, on a single Nvidia 2080Ti GPU.
% During training, we randomly cropped the audio clips to 7960 samples and set the batch size to 16.
% The model is optimized using the AdamW optimizer \cite{loshchilov2018decoupled} with $\beta_1=0.8$, $\beta_2 = 0.99$. The learning rate was set initially to $2 \times 10^{-4}$ and scheduled to decay with a factor of 0.999 at every epoch.

\subsubsection{Baselines.} We compared ESTVocoder with HiFi-GAN \cite{kong2020hifi} \footnote{\href{https://github.com/jik876/hifi-gan}{https://github.com/jik876/hifi-gan}.}, SiFi-GAN \cite{yoneyama2023source} \footnote{\href{https://github.com/chomeyama/SiFi-GAN}{https://github.com/chomeyama/SiFi-GAN}.} and Vocos \cite{siuzdak2023vocos} \footnote{\href{https://github.com/gemelo-ai/vocos}{https://github.com/gemelo-ai/vocos}.}. 
SiFi-GAN introduces excitation into HiFi-GAN, constructing a neural vocoder within the source-filter framework. 
In contrast to HiFi-GAN and SiFi-GAN, Vocos models the amplitude and phase spectra instead of directly modeling the waveform. 
However, it is not designed within the source-filter framework. 
Therefore, comparing them with the proposed ESTVocoder is fair and reasonable.
All the baseline models were trained on the same LJSpeech dataset with the same training environment as ESTVocoder.

%a batch size of 16 for 200,000 steps on a single NVIDIA RTX TITAN, which is a relatively small number of steps for a vocoder.
\vspace{-1.5mm}
\subsubsection{Tasks.} 
\vspace{-1.5mm}
In this paper, we conducted analysis-synthesis experiments on HiFi-GAN, SiFi-GAN, Vocos, and ESTVocoder. 
We performed TTS experiments only on HiFi-GAN, Vocos, and ESTVocoder using mel spectrograms predicted by FastSpeech2 \cite{ren2020fastspeech} \footnote{\href{https://github.com/ming024/FastSpeech2}{https://github.com/ming024/FastSpeech2}.}. 
Since SiFi-GAN utilized mel cepstral features extracted by WORLD \cite{morise2016world} and no TTS experiments were conducted in the original paper \cite{yoneyama2023source}, we also discarded the TTS experiment on SiFi-GAN.

\vspace{-1.5mm}
\subsubsection{Evaluation metircs.}
\vspace{-1.5mm}
For the analysis-synthesis experiment, we have adopted a variety of objective metrics to evaluate the quality of synthesized speech from multiple perspectives.
To evaluate the amplitude-related quality, the root mean square error of logarithmic amplitude spectrum (LAS-RMSE) and mel cepstrum distortion (MCD) were used. 
To evaluate the F0 modeling accuracy, root mean square error of F0 (F0-RMSE) and V/UV flag error (denoted by V/UV) were adopted. 
Finally, to comprehensively evaluate the quality of synthesized speech, we introduced two commonly used tools, including perceptual evaluation of speech quality (PESQ) \cite{rec2007p} and virtual speech quality objective listener (ViSQOL) \cite{chinen2020visqol}.

% including perceptual evaluation of speech quality (PESQ), virtual Speech Quality Objective Listener (ViSQOL) \cite{chinen2020visqol}, root mean square error of F0 (F0-RMSE), root mean square error of logarithmic amplitude spectrum (LAS-RMSE), mel cepstrum distortion (MCD), and V/UV error.

To evaluate the generation efficiency of vocoders, we compared the real-time factor (RTF) and training speed. 
The RTF is calculated as the ratio of generation time and the actual duration of the test set. 
The training speed is defined as the seconds taken to complete one epoch of training (s/e), which reflects the efficiency of model training.

For the TTS experiment, we employed a subjective mean opinion score (MOS) test, to compare the naturalness of the vocoders. 
In each MOS test, 20 test utterances synthesized by compared vocoders along with the natural utterances were compared and evaluated by at least 30 native English listeners. 
This test was conducted on the Amazon Mechanical Turk crowd-sourcing platform, where listeners were asked to rate the naturalness on a scale from 1 to 5, with a score interval of 0.5.

\begin{table}[t]
	\centering
	\caption{Quality-related objective evaluation results of for ESTVocoder, HiFi-GAN, SiFi-GAN and Vocos on the test set of the LJSpeech for the analysis-synthesis experiments. The \textbf{bold} and \underline{underline} numbers indicate the optimal and sub-optimal results, respectively.}\label{tab1}
	\adjustbox{width=0.8\textwidth}{
		\begin{tabular}{l c c c c c c}
			\hline
			\hline
			\multirow{2}{*}  &{\scriptsize {LAS-RMSE}
} &{\scriptsize{MCD}
} &\scriptsize {F0-RMSE} &{\scriptsize{V/UV}
}& {\scriptsize PESQ$\uparrow$}& {\scriptsize ViSQOL$\uparrow$}\\  &{ {\scriptsize(dB)$\downarrow$}} &{\scriptsize (dB)$\downarrow$} &{ {\scriptsize(cent)$\downarrow$}} &{ {\scriptsize (\%)$\downarrow$}}&  {  {\scriptsize} }& { {\scriptsize}}\\
			\hline
			{\scriptsize ESTVocoder}  &\underline{36.51} &\textbf{0.931} &\textbf{4.17} &\textbf{4.99}& \textbf{3.70}&\textbf{4.895}  \\
			%\hline
			{\scriptsize HiFi-GAN}  &40.34 &1.227 &5.07 &5.23&3.48&4.815 \\
   
			%\hline
			{\scriptsize SiFi-GAN}  &\textbf{29.39} &1.252 &5.03 &\underline{5.15}& 3.33&4.703\\
  		{\scriptsize Vocos}  &39.07 &\underline{1.033} &\underline{4.40} &5.31& \underline{3.51}&\underline{4.863} \\ 
			%\hline
			
			\hline
			\hline
	\end{tabular}}
\end{table}

\begin{table}[t]
	\centering
	\caption{Efficiency-related objective evaluation results of for ESTVocoder, HiFi-GAN, SiFi-GAN and Vocos on the test set of the LJSpeech for the analysis-synthesis experiments. The \textbf{bold} and \underline{underline} numbers indicate the optimal and sub-optimal results, respectively. Here, s/e represents second per epoch and "$a\times$" represents $a\times$ real time.}\label{tab2}
	\adjustbox{width=0.8\textwidth}{
		\begin{tabular}{l c c c}
			\hline
			\hline
			\multirow{2}{*} &  {\scriptsize RTF
}& {\scriptsize RTF
}& {\scriptsize Training Speed
}\\  &{\scriptsize (GPU)$\downarrow$}&{\scriptsize (CPU)$\downarrow$}&    {\scriptsize(s/e)$\downarrow$}\\
			\hline
            
			{\scriptsize ESTVocoder} & 0.0016 (625.0$\times$)&\underline{0.0112 (89.3$\times$)}&\underline{203} \\
			%\hline
			{\scriptsize HiFi-GAN} &\underline{0.0013 (769.2$\times$)}&0.1496 (6.8$\times$)&358\\
   
			%\hline
			{\scriptsize SiFi-GAN} & 0.0135 (74.1$\times$)&0.1568 (6.4$\times$)&567\\
  			{\scriptsize Vocos} & \textbf{0.0009 (1111.1$\times$)}&\textbf{0.0093 (107.5$\times$)}&\textbf{199}\\ 
			%\hline
			
			\hline
			\hline
	\end{tabular}}
\end{table}

\vspace{-1.5mm}
\subsection[Evaluation Results]{Evaluation Results\footnote{Audio samples of the proposed ESTVocoder can be accessed at \href{https://pb20000090.github.io/NCMMSC2024/}{https://pb20000090.github.io/NCMMSC2024/}.}}
\vspace{-1.5mm}
% \subsection{Evaluation Results}\footnote{Audio samples of the proposed ESTVocoder can be accessed at\href{https://pb20000090.github.io/NCMMSC2024/}{https://pb20000090.github.io/NCMMSC2024/}.}
\label{ssec:eva}
The quality-related objective evaluation results of the proposed ESTVocoder and baseline HiFi-GAN, SiFi-GAN and Vocos for the analysis-synthesis experiments are presented in Table \ref{tab1}. 
Regarding the amplitude quality, it can be observed that our proposed ESTVocoder achieved the lowest MCD results, but it was second only to SiFi-GAN in the LAS-RMSE metric. 
However, the MCD calculation is performed in the mel domain, which better reflects human perception. 
Therefore, it can be inferred that the synthesized speech of our proposed ESTVocoder had higher perceived quality. 
For F0 recovery capability, our proposed ESTVocoder had the lowest F0-RMSE and V/UV error, indicating its strong F0 restoration capability and fewer pronunciation errors. 
Thanks to ESTVocoder's excellent performance in amplitude spectrum perceptual quality and fundamental frequency recovery capability, it is unsurprising that its overall speech quality is also the best, according to PESQ and ViSQOL results. 
Therefore, our proposed ESTVocoder outperformed commonly used waveform-prediction-based neural vocoders (i.e., HiFi-GAN), excitation-waveform-transformed neural vocoders (e.g., SiFi-GAN), and spectral-prediction-based neural vocoders (i.e., Vocos) in terms of synthesized speech quality. 
This confirms the advantages of our proposed excitation spectral transformation method. 

%The quality evaluation comparison of the synthesized speech between our proposed ESTVocoder and other vocoders is presented in Table \ref{tab1}. 
% It can be observed that our model outperforms others in terms of PESQ, ViSQOL, F0-RMSE, V/UV error and MCD, indicating that our newly introduced excitation signal is beneficial for improving the quality of synthesized speech. Additionally, in terms of the LAS-RMSE, our model ranks second only to SiFi-GAN and outperforms other models. This indicates that the inclusion of an excitation signal based on the F0 facilitates the model's ability to learn certain acoustic details more easily. However, SiFi-GAN does not perform well on other metrics compared to other vocoders.

Table \ref{tab2} gives the results of the comparison of the generation and training speeds among the proposed ESTVocoder, HiFi-GAN, SiFi-GAN and Vocos for the analysis-synthesis experiments. 
We first compared the proposed ESTVocoder with Vocos, as both were designed to predict spectra rather than waveforms. 
The generation and training efficiency of ESTVocoder were slightly lower than those of Vocos, which is reasonable given that ESTVocoder introduced an additional excitation producer module. 
This conclusion is consistent with the comparison between HiFi-GAN and SiFi-GAN. 
When compared with the two waveform-prediction-based neural vocoders, i.e., HiFi-GAN and SiFi-GAN, although our proposed ESTVocoder had a slightly slower generation speed on GPU compared to HiFi-GAN, its generation speed on CPU was significantly faster than both HiFi-GAN and SiFi-GAN. 
This demonstrates the efficiency advantage of predicting spectra rather than waveforms, especially when GPU parallel acceleration is not available. 
Additionally, the training efficiency of ESTVocoder is also significantly higher than that of HiFi-GAN and SiFi-GAN.

% It can be seen that on GPU, despite the introduction of a new excitation signal, ESTVocoder does not significantly fall behind Vocos or HiFi-GAN in terms of generation speed. On CPU, the generation speed of ESTVocoder remains as fast as that of Vocos and significantly surpasses HiFi-GAN. ESTVocoder also retains the high training speed characteristic of Vocos. As for SiFi-GAN, it lags behind the other vocoders in all three of these metrics.

\begin{table}[t]
	\centering
	\caption{Subjective evaluation results of ESTVocoder, HiFi-GAN and Vocos on the test set of the LJSpeech for the TTS experiments.}\label{tab1MOS}
	\adjustbox{width=0.3\textwidth}{
		\begin{tabular}{l c}
			\hline
			\hline
			& {\scriptsize{MOS$\uparrow$}} \\
			\hline
            {\scriptsize Natural}  &4.01±0.044  \\
			{\scriptsize ESTVocoder}  &3.87±0.051  \\
			%\hline
			{\scriptsize HiFi-GAN}  &3.90±0.047 \\
  			{\scriptsize Vocos}  &3.87±0.049 \\ 
   
			%\hline
			
			\hline
			\hline
	\end{tabular}}
\end{table}

For TTS experiments, the results of the subjective MOS test on the natural speech and speeches generated by ESTVocoder, HiFi-GAN and Vocos are shown in Table \ref{tab1MOS}. 
We can see that their average MOS was similar. 
To determine the significance of the differences among them, we also calculated the $p$-value of the $t$-test. 
The results of the two $t$-tests between ESTVocoder and HiFi-GAN and between ESTVocoder and Vocos indicated that the performance differences between the ESTVocoder and other baseline vocoders were not significant ($p$ > 0.05) in the TTS task. 
Therefore, our proposed ESTVocoder is comparable to other baseline vocoders in terms of subjective quality on TTS tasks. 
However, it achieves better objective results in analysis-synthesis tasks and has considerable training and generation speed advantages. 
In the following subsection, we also demonstrate that ESTVocoder has a faster training convergence speed, thus saving training costs, which is also one of its advantages.

% It can be seen from Table \ref{tab1} that in the TTS task, the quality difference between the speech synthesized by ESTVocoder and Vocos is not significant, both are slightly inferior to HiFi-GAN, and overall, they are comparable.

% \begin{table}
% 	\centering
% 	\caption{Subjective evaluation results of TTS experiment. The width of the 95\% confidence interval is given in parentheses.}\label{tab3}
% 	\adjustbox{width=0.9\textwidth}{
% 		\begin{tabular}{l c c c c}
% 			\hline
% 			\hline
% 			 &  {Ground Truth
% }&  {ESTVocoder
% }& {Vocos
% }& {HiFi-GAN
% }\\  
% 			\hline
            
% 			{MOS} &4.01(±0.044)&3.87(±0.051)&3.87(±0.049)&3.90(±0.047) \\

% 			\hline
% 			\hline
% 	\end{tabular}}
% \end{table}

% Besides the Mean Opinion Score (MOS), we also utilized the $p$-value from the $t$-test to assess the significance of the differences between the vocoders. The results of the three sets of $t$-tests between ESTVocoder, Vocos, and HiFi-GAN indicate that the performance differences among them are not significant ($p$ > 0.05) in the TTS task.

\begin{figure}[t]
	\centering
	\includegraphics[width=0.8\textwidth]{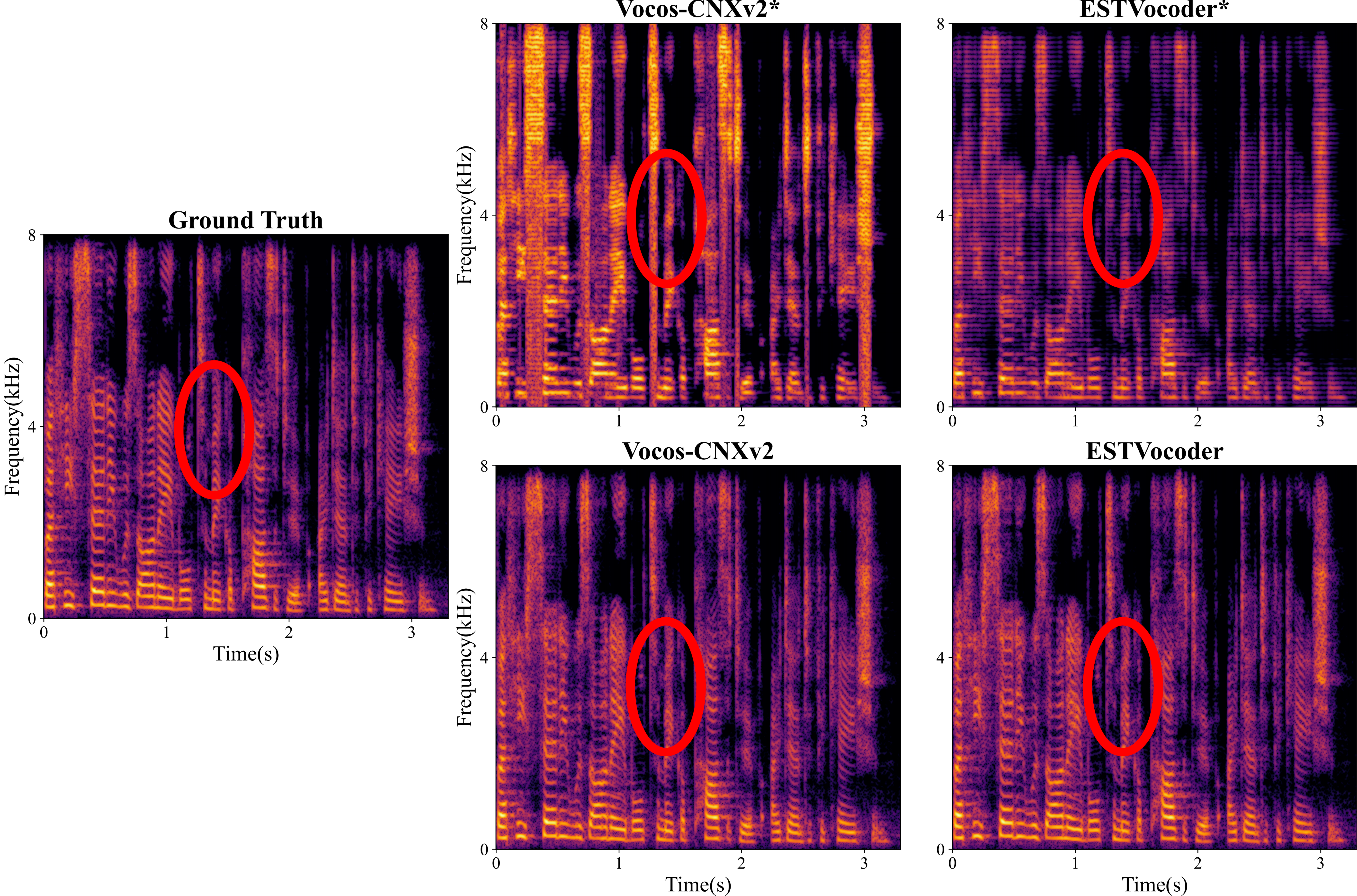}
	\caption{Visualized amplitude spectra of natural speech and speeches generated by Vocos-CNXv2*, ESTVocoder*, Vocos-CNXv2 and ESTVocoder (example sentence LJ034-0143)} \label{analysis}
\end{figure}

\subsection{Analysis and Discussion}

\subsubsection{Discussion on backbone.} 
Choosing a good backbone is crucial for the performance of the synthesized speech by a neural vocoder. 
As mentioned in Section \ref{ssec:block}, we choose ConvNeXt v2 blocks as the backbone for our proposed ESTVocoder due to their proven strong modeling capabilities. 
However, Vocos uses ConvNeXt blocks as its backbone. 

\begin{table}[t]
	\centering
	\caption{Quality-related objective evaluation results of for ESTVocoder, ESTVocoder-CNXv1, Vocos-CNXv2 and Vocos on the test set of the LJSpeech for the analysis-synthesis experiments.}\label{tab4}
	\adjustbox{width=0.8\textwidth}{
		\begin{tabular}{l ccccc c }
			\hline
			\hline
			\multirow{2}{*}  &{\scriptsize {LAS-RMSE}
} &{\scriptsize{MCD}
} &\scriptsize {F0-RMSE} &{\scriptsize{V/UV}
}& {\scriptsize PESQ$\uparrow$}& {\scriptsize ViSQOL$\uparrow$}\\  &{ {\scriptsize(Hz)$\downarrow$}} &{\scriptsize (dB)$\downarrow$} &{ {\scriptsize(dB)$\downarrow$}} &{ {\scriptsize (\%)$\downarrow$}}&  {  {\scriptsize} }& { {\scriptsize}}\\
			\hline
			{\scriptsize ESTVocoder}  &\textbf{36.51} &\textbf{0.931} &\textbf{4.17} &\textbf{4.99}& \textbf{3.70}&\textbf{4.895}\\
			%\hline

			{\scriptsize ESTVocoder-CNXv1}&37.60 &1.025 &4.27 &5.46&3.63&4.880\\
   
			%\hline
			{\scriptsize Vocos-CNXv2}&38.53 &0.998 &4.30 &5.30& 3.58&4.878\\
			  			{\scriptsize Vocos}  &39.07 &1.033 &4.40 &5.31& 3.51&4.863\\ 
			
			\hline
			\hline
	\end{tabular}}
\end{table}

Although researchers have systematically compared the differences between ConvNeXt and ConvNeXt v2 in the field of image processing \cite{woo2023convnext}, this difference has yet to be systematically validated in the field of speech signal processing. 
Therefore, we replaced all ConvNeXt v2 blocks in ESTVocoder with ConvNeXt blocks (referred to as ESTVocoder-CNXv1) and compared it with the original ESTVocoder. 
Simultaneously, we replaced all ConvNeXt blocks in Vocos with ConvNeXt v2 blocks (referred to as Vocos-CNXv2) and compared it with the original Vocos. 
Objective experimental results are listed in Table \ref{tab4}. 
We can observe that in all metrics, the ESTVocoder outperformed the ESTVocoder-CNXv1, and the Vocos-CNXv2 outperformed the Vocos. 
This result indicates that ConvNeXt v2 has a stronger modeling capability in vocoder tasks compared to ConvNeXt, which is why we chose ConvNeXt v2 as the backbone for ESTVocoder.

\begin{figure}[t]
	\centering
	\includegraphics[width=7cm]{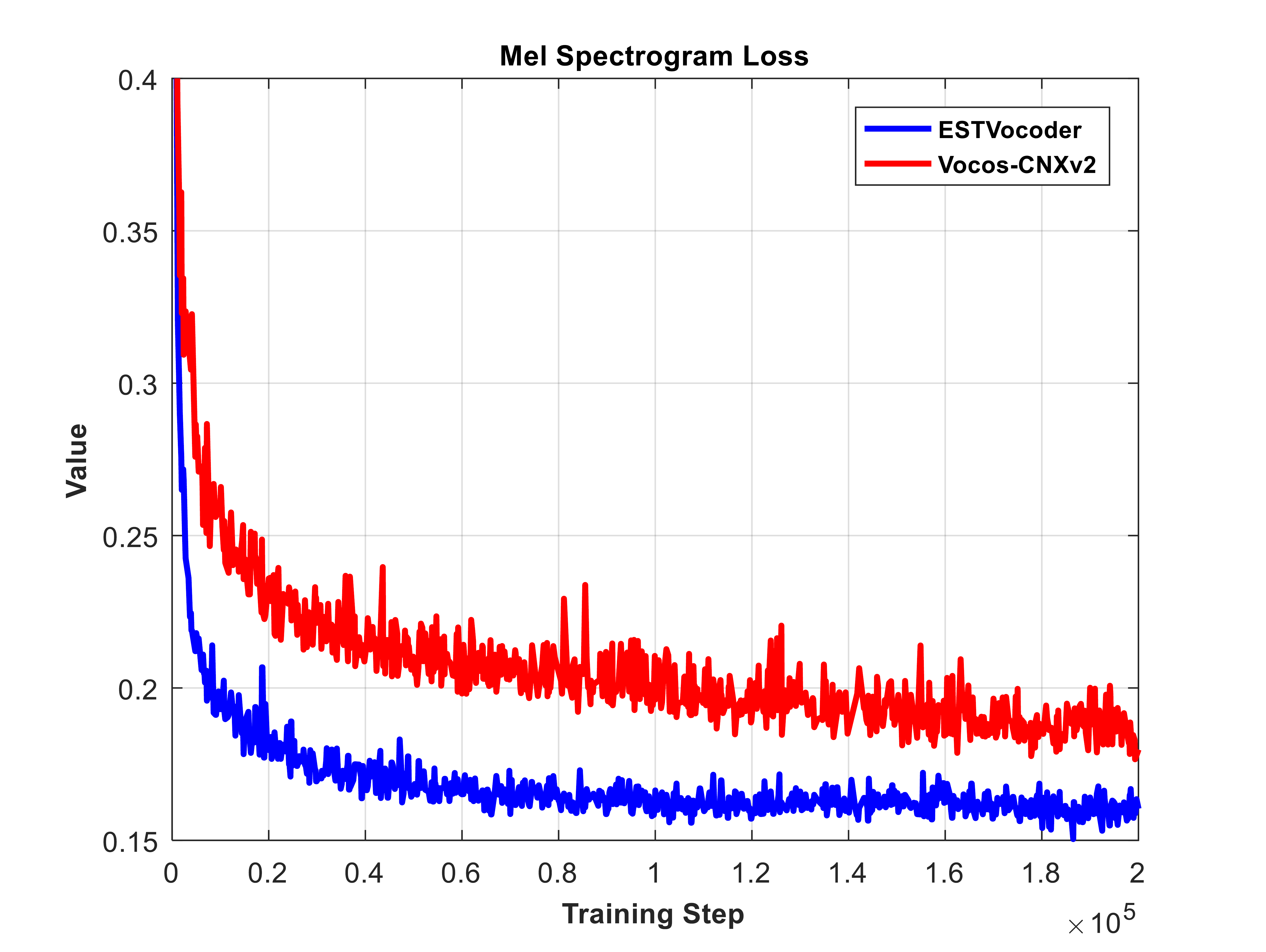}
	\caption{Training curves of mel-spectrogram loss for ESTVocoder and Vocos-CNXv2.} \label{analysis2}
\end{figure}

\vspace{-1.5mm}
\subsubsection{Analysis of the role of the excitation producer.} 
\vspace{-1.5mm}
As mentioned in Section \ref{sec:excit}, the original intention of introducing the excitation spectra in ESTVocoder is to provide the neural filter with prior information about the speech spectra, thereby reducing the training difficulty of the neural filter and improving the quality of the synthesized speech. 
To confirm this point, we compared ESTVocoder with Vocos-CNXv2. 
They both use ConvNeXt v2 as the backbone, so Vocos-CNXv2 can be approximately regarded as the ablated version of ESTVocoder without the excitation producer. 
We conducted the following three analyses. 
1) Objective evaluation. As shown in Table \ref{tab4}, the ESTVocoder outperformed the Vocos-CNXv2 for all objective metrics. 
This confirms that the introduced excitation producer significantly helps improve the quality of synthesized speech. 
2) Visual analysis. Both ESTVocoder and Vocos-CNXv2 used 8 ConvNeXt v2 blocks, so we attempted to reconstruct the speech waveform based on the output from their 4th block (referred to as ESTVocoder* and Vocos-CNXv2*, respectively) to test their performance at intermediate stages of the model. 
Figure \ref{analysis} shows the amplitude spectra of the speeches generated by Vocos-CNXv2*, ESTVocoder*, Vocos-CNXv2, and ESTVocoder. 
Interestingly, the amplitude spectrum of ESTVocoder* is close to that of natural speech, while the amplitude spectrum of Vocos-CNXv2* is significantly worse. 
This indicates that the speech spectral prior introduced by the excitation allows the neural filter to quickly learn the shape of the speech spectra, resulting in a faster learning speed. 
By comparing Vocos-CNXv2 and ESTVocoder, as shown by the red circles in Figure \ref{analysis}, ESTVocoder's harmonic restoration is more accurate, which is also attributed to the introduction of the excitation prior information. 
3) Convergence speed analysis. Figure \ref{analysis2} plots the training curves of the mel spectrogram loss for Vocos-CNXv2 and ESTVocoder. 
From the figure, we can also see that ESTVocoder's convergence speed is significantly faster, which is also attributed to the excitation prior information provided to the neural filter.

% In this section, we will analyze how the newly introduced excitation signal in ESTVocoder can more rapidly reduce the training loss and enhance the quality of synthesized speech.

% To explore the specific impacts of the different modifications we made to Vocos on analysis-synthesis experiment: upgrading the model structure to ConvNeXt v2 and introducing an excitation signal, we conducted comparative experiments. In these comparative experiments, we trained versions of ESTVocoder using ConvNeXt v1 (ESTVocoder-v1) and Vocos using ConvNeXt v2 (Vocos-v2), and compared them with ESTVocoder and Vocos.

% From Table \ref{tab4}, it can be observed that after replacing with the ConvNeXt module, all metrics of Vocos and ESTVocoder have improved, indicating that the modeling capability of ConvNeXt v2 is stronger than the ConvNeXt block. With the same block, adding the excitation signal as input also significantly improves all metrics, demonstrating the effectiveness of the strategy adopted by our proposed ESTVocoder.

% To demonstrate the effect of the excitation signal, we reduced the 8-layer residual network of ESTVocoder and Vocos-v2 to only 4-layer, leaving the rest unchanged, and then conducted speech synthesis.

% From the Fig. \ref{analysis}, it can be observed that the amplitude spectrum of the audio synthesized by ESTVocoder is not as smooth as that of Vocos-v2, indicating that the introduced excitation signal serves to complement the amplitude spectrum and prevent the introduction of artificial artifacts.

\vspace{-1.5mm}
\section{Conclusions}
\vspace{-1.5mm}
\label{sec:con}

In this paper, we present an excitation-spectral-transformed neural vocoder called ESTVocoder under the source-filter framework. 
An excitation producer produces the excitation waveform according to F0, and then the neural filter transforms the excitation amplitude and phase spectra to speech ones conditioned on the mel spectrogram. 
The comprehensive experimental results demonstrate that our proposed ESTVocoder outperforms several baseline vocoders. 
Analysis experiments confirm that the excitation provides the neural filter with speech spectral priors, effectively reducing the training difficulty of the neural filter and improving the quality of synthesized speech. 
Further improving the efficiency of ESTVocoder and applying it to other tasks (i.e., speech enhancement) will be the focus of our future work.

% In this paper, we introduce ESTVocoder. The input to ESTVocoder is an excitation signal based on the fundamental frequency (F0) that contains all possible harmonics. Guided by the mel spectrogram as a condition, the excitation signal is directed to generate the final waveform. ESTVocoder significantly enhances the performance of analysis-synthesis tasks while maintaining the high efficiency and speed of Vocos. Subjective experiments have demonstrated that in TTS tasks, it is comparable to HiFi-GAN and Vocos. Moreover, analysis experiments have confirmed that the excitation signals we introduced, as well as the ConvNeXt v2 modules, are effective. In the future, we will focus on how to apply ESTVocoder to other speech synthesis tasks, such as Speech Enhancement and Voice Conversion.
\bibliographystyle{splncs04_unsort}
\bibliography{mybibliography}

\end{document}